\documentclass[12pt]{article}
\usepackage{graphicx} 
\usepackage{amssymb}
\usepackage{graphics}
\usepackage{amsmath}
\usepackage{multirow}
\usepackage{bbm}
\usepackage{graphicx}
\usepackage{enumitem}
\usepackage{array}   
\usepackage{siunitx,booktabs}
\usepackage{float}
\usepackage[left=1in, right=1in, top=1in, bottom=1in]{geometry}
\graphicspath{ {./images/} }

\title{Statistical Design and Planning of an Adaptive Trial using Hierarchical Composite Outcomes: A Practical example}
\author{Krishna Padmanabhan, Cyrus Mehta}
\date{Jan 2025}

\begin{document}
\maketitle

\section*{Introduction}
\subsection*{Composite endpoints}
Selecting appropriate endpoints is crucial for evaluating treatment efficacy in randomized controlled trials (RCTs). Composite endpoints enable smaller sample sizes and shorter trials, reducing costs. However, they present challenges in interpretation, particularly when components vary in clinical significance. The traditional first-event analysis ignores subsequent outcomes, potentially undervaluing serious events like death.

\subsection*{The Finkelstein-Schoenfeld (FS) statistic}
\noindent The Finkelstein-Schoenfeld (FS) test has gained prominence in cardiovascular trials as a method for analyzing prioritized composite endpoints. This approach allows researchers to hierarchically arrange multiple outcomes based on their clinical importance, with mortality typically given the highest priority. The FS test conducts pairwise comparisons between patients, assigning scores based on their performance across the hierarchy of endpoints, while maintaining the relative clinical relevance.

\subsection*{Advantages of Sample Size Re-estimation (SSR)}
When, treatment effect and endpoint variance estimates are uncertain, sample sizes can be hard to calculate. Such uncertainty can result in trials being underpowered, risking failure, or overpowered, leading to unnecessary expense and scale. In these scenarios, the ability to adjust sample size based on interim data is highly beneficial. Sample size re-estimation (SSR) involves reassessing the sample size during interim analysis, conducted midway through a trial on a portion of the planned sample. This belongs to a class of well-understood designs from both a statistical perspective and regulatory acceptability.  

\subsection*{SSR and F-S statistic}
Adaptive design methodology for complex or hierarchical endpoints, such as those analyzed using the F-S statistics remains underdeveloped. This paper aims to introduce a statistical framework for sample size re-estimation during interim analysis, leveraging approximations when closed-form solutions are unavailable.

\section*{Proposed Design}
\subsection*{Study Background}
This study, inspired by a real-world trial where the authors contributed to the statistical design and implementation, evaluates the effect of an intervention on cardiovascular disease. The trial is a two-arm adaptive clinical study, with participants randomly and equally allocated to either the treatment arm (``Active") or the Control arm in which standard therapy is administered.
\newline
\newline
\noindent In this study, the primary endpoint is a hierarchical composite outcome, incorporating all-cause mortality, the frequency of CV-related hospitalizations (CVH), and improvements in patient functioning, assessed via a physical test, similar to the 6-minute walk test (6MWT). The evaluation follows a predefined hierarchical order:
\begin{enumerate}
    \item The time until the occurrence of  all-cause death
    \item The count of CVH (adjudicated by a blinded committee)
    \item Achieving a response on the functional test at the 12 month evaluation (0 or 1, 1 indicating a positive response).
\end{enumerate}

\subsection*{F-S Statistic Calculation and Hypothesis Test}
The F-S statistic is calculated by comparing each subject $i$ to every other subject $j$ from the data set and designating the results of this comparison with the rank $U_{ij}$ in accordance with the below hierarchical ranking algorithm:
\newline
\newline 
\noindent For each pair of subjects $i$ and $j$, define a score $u_{ij}$ with these steps:
\begin{enumerate}
    \item If subject $i$ is known to have experienced death at a later date than subject $j$, then $u_{ij}$ = 1. Similarly, if subject $j$ is known to have experienced death at a later date than subject $i$, then $u_{ij}$ = -1. 
    \item If it is not known which subject has gone longer without death, then compare the number of CVH for the two subjects within the shorter of the two follow-up durations. And assign a rank of 1 or -1 similarly. 
    \item If a win score cannot be determined based on either endpoint, then compare the response rates for the two subjects on the functional assessment and assign a rank of 1 or -1 similarly. 
    \item For any pair of subjects where ties cannot be broken despite evaluation of all three endpoints, $u_{ij} = u_{ji}$ = 0, and in all cases, $u_{ij} = -u_{ji}.$ \\ \\
\noindent For further details on the calculation, refer to Finkelstein and Schoenfeld (1999). The authors also extended their initial research to design of longitudinal studies with similar endpoints in their 2022 (Schoenfeld et al) publication. 
\end{enumerate}
Suppose $n$ subjects are randomized to each arm with total $2n$ subjects being randomized in the trial. Each
subject $i$ is assigned a score:
\begin{equation}
    U_i = \sum_{\substack{j = 1 \\ j \neq i}}^{2n} U_{ij}.
\end{equation}
\noindent Let $Z_i$ be a treatment indicator that takes a value of 1 if subject $i$ is randomized to the Device arm, and 0 if subject $i$ is randomized to the Control arm. The F-S statistic can then be written as: 
\begin{equation}
    T = \sum_{i=1}^{2n} Z_iU_i.
\end{equation}
This statistic is asymptotically normal with mean:
\begin{equation}
    E(T) = 2n^2(\theta - 1/2),
\end{equation}
where $\theta$ is the probability that a random subject $i$ in the treatment group has a better outcome than a random subject $j$ in the control group. Assuming ties exist, the variance of the test statistic under the null is: 
\begin{equation}
    Var(T) = \frac{n^2}{2n(2n-1)}\sum_{i=1}^{2n} U_i^2.
\end{equation}

\subsection*{Study Hypotheses}

\noindent The null hypothesis of no treatment effect with respect to death, number of CV hospitalizations, and improvement in function  can then be expressed as: 
\[
H_0 : \theta = \frac{1}{2}
\]
and the one-sided alternative hypothesis is 
\[
H_A : \theta > \frac{1}{2}.
\]
The hypothesis test will be conducted assuming a one-sided Type-I error rate of 2.5$\%$. 

\subsection*{Initial Sample Size and Interim Details}
The  study will employ a 2-stage design. A total of 400 randomized subjects are planned to be enrolled, assuming no sample size increase at the interim analysis. The unblinded interim analysis will be conducted on the data from the first 200 subjects, who have each been followed for 12 months, unless censored earlier by death or drop-out. The purpose of this interim analysis is two-fold: to conduct a futility analysis (non-binding) and to determine whether a sample size increase is needed. The final analysis for the primary outcome will be conducted when all subjects not censored by death or drop-out have each been followed for 12 months. The rates of death, hospitalization and response on the functional test for the null and alternative hypothesis were set based on historical data and clinically meaningful changes. These values were as follows: 
\begin{itemize}
\item {Prob. of Death: 30$\%$ Active vs. 40$\%$ control}
\item {CVH Hazard Rates: 0.25 Active vs. 0.4 control}
\item {Repsonse rate on Function scale: 50$\%$ Active vs. 25$\%$ control}
\end{itemize}

\noindent Based on these rates, data were simulated for 10,000 hypothetical trials to establish an initial sample size. Utilizing the methodology of Yu and Ganju (2023), N = 400 subjects would be sufficient for approximately 90$\%$ power. This SS calculation method is based on the Win Ratio Statistic, which was developed by Pocock et. al. (2012). The Win Ratio is an extension of the FS statistic, providing an intuitive and interpretable estimate of the treatment effect in trials using a hierarchical composite endpoint. From the simulated data, estimates for the observed Win Ratio (approx. 1.52) under the alternative hypothesis, and its variance (approx. 6.8) were calculated. The empirical probability of a tie was estimated to be around 12.5$\%$.

\subsection*{Interim Analysis}
The interim analysis serves to establish whether the trial should stop for futility (non-binding) an whether an increase in the sample size is required. The decision rule for SSR is based on a partition of the sample space, based on the calculated predictive probability (PP). The PP is defined as the probability that $H_0$ will be rejected when the analysis is performed when the planned N patients have completed the trial, conditional on the observed data. \\

\noindent Based on simulation studies, the following partition of the Interm Analysis PP space was decided upon. If the estimated PP falls under 10$\%$, then our trial is terminated at the interim analysis, and futility is declared. If the PP is between 10-30$\%$ or  $\ge$ 90$\%$, the trial will continue as planned to the pre-planned N of 400 subjects. However, if the estimated PP falls in a 'promising' zone, defined as a PP between (30$\%$, 90$\%$) the total sample size will increase. If the PP at the IA falls between (30$\%$, 75$\%$), then the Sample Size for the full trial will be increased to 600 patients (i.e., 400 additional patients in Stage 2). And if the PP is between (75$\%$, 90$\%$), then the Sample Size for the full trial will be increased to 500 patients (i.e., an additional 100 patients over the plan). These zones are summarized in Table 1.

\subsection*{Estimation of Predictive Power at IA}
We use the approximation from Marion et. al. in order to calculate the predictive probability of final analysis success for our two stage trial. At the interim analysis, consider a normal test statistic $Z_n$ with $p$-value $p_n$. The predictive probability $PP_N$ measures the likelihood of rejecting the null hypothesis when $N$ (pre-planned SS) patients complete the trial. At the interim analysis, we know the future information level $I_N$, but both the final statistic $Z_N$ and $p$-value $p_N$ remain unknown. Using the interim $p$-value $p_n$ and information fraction $r = I_n/I_N$ where $0 < I_n < I_N$, we can approximate $PP_N$ as:

\begin{equation}
PP(p_n, r, \alpha) = \Phi\left(\frac{\Phi^{-1}(1-p_n) - \Phi^{-1}(1-\alpha)\sqrt{r}}{\sqrt{1-r}}\right)
\end{equation}

\subsection*{Stage 2 Sample Size Re-Estimation}

In our study, the zones for futility stop or adaptive sample size increase based on PP were determined and optimized based on simulations and are listed below. 
\begin{table}[h]
\caption{Interim Decision Table}
\begin{tabular}{ |p{3.5cm}|p{4.5cm}|p{4.5cm}|}
\hline
Zone & Partition & Decision \\
\hline
  Futility & $\text{PP}(t_1, n_2) < 0.1$ & Stop trial for futility \\
  Unfavorable & $0.1 \leq \text{PP}(t_1, n_2) < 0.3$ & Cohort 2 SS = 200 \\
  Promising-Low & $0.3 \leq \text{PP}(t_1, n_2) < 0.75$ & Cohort 2 SS = 400 \\
Promising\_High & $0.75 \leq \text{PP}(t_1, n_2) < 0.9$ & Cohort 2 SS = 300\\  
Favorable & $\text{PP}(t_1, n_2) \geq 0.9$ & Cohort 2 SS = 200\\

\hline
\end{tabular}
\end{table}
\newline
 
\subsection*{Final Analysis}
Once the predictive power has been estimated and the sample size for stage 2 has been re-evaluated, stage 2 data will be collected and final analysis will be conducted once all subjects in the study have been followed for 12 months (or terminated early). The decision to reject the null hypothesis is based on the weighted combination of the test statistics from stages 1 and 2, with a rejection occurring when: 
\begin{equation}
    w_1\frac{t_1}{s_1} + w_2\frac{t_2}{s_2} \geq z_\alpha, \  \text{for the case of no SSR} 
\end{equation}
\begin{equation}
 w_1\frac{t_1}{s_1} + w_2\frac{t_2^{*}}{s_2^{*}} \geq z_\alpha, \  \text{for the case of SSR},
\end{equation}
where each $(t_i, s_i)$ is evaluated by equations (2) and (4), from the $n_i$ subjects belonging to cohort $i$. It is important to note that the weights, $w_1 = \sqrt{\frac{n_1}{n}}$ and $w_2 = \sqrt{\frac{n_2}{n}}$ are determined by the pre-declared sample sizes, and thus are the same regardless of whether or not SSR occurs. This is equivalent to the Cui-Hung-Wang (CHW) method, which preserves the type I error, which may be inflated due to the data-dependentd SSR.\\

\section*{Simulation}
In this section we outline the process of simulating this entire trial repeatedly under various parameter assumptions. Table 1 displays five hypothetical scenarios of treatment response versus control arm response for mortality hazard rates under exponential survival, CVH rates under Poisson arrivals, and probabilities of response on the functional scale (binomial).
\begin{center}
\begin{table}[h]
\caption{Hypothetical Scenarios to be Explored}
\makebox[\textwidth]{
\begin{tabular}{ |p{2.2cm}|p{2.18cm}|p{2.35cm}|p{2.18cm}|p{2.5cm}|p{2.18cm}|}
 \hline
 \multicolumn{6}{|c|}{Five Hypothetical Scenarios (Active vs. Control)} \\
 \hline
 Event Rates & Null & Alternative &  Alternative2 & Middling & Middling2 \\
 \hline
 Mortality   & 40\% vs 40\% & 30\% vs 40\% &   40\% vs 40\% & 32.5\% vs 40\% & 35\% vs 40\% \\
 CVH Rate &   0.4 vs 0.4  & 0.25 vs 0.375 & 0.10 vs 0.40 & 0.275 vs 0.40 & 0.28 vs 0.4 \\
 Func. Resp. & 25\% vs 25\% & 50\% vs 25\% &  60\% vs 25\% & 45\% vs 30\% & 40\% vs 25\%\\
 \hline
\end{tabular}}
\end{table}
\end{center}

\noindent The scenarios systematically evaluate statistical properties across varying treatment effects. The null scenario intends to verify type I error control. Two alternative scenarios represent optimistic effects mediated to varying degrees by different endpoints, while the two ``middling'' scenarios model modest effects. These scenarios are particularly valuable for evaluating sample size re-estimation procedures, as they represent treatment effects at the lower boundary of clinical significance, where maintaining adequate power may still be useful.\\

\subsection*{Intra-patient Data Model}
\noindent As previously mentioned, patient level data for each subject is generated according to a joint frailty model and the treatment effects specified in Table 2. This joint-frailty model is employed to account for the correlation between a subject's time to death and CV hospitalizations. We largely follow the model specified in Rogers (2014). The Joint Frailty Model (JFM) captures risk of death from repeated CV hospitalizations while handling competing death risks. The model conditions these outcomes on a frailty term.\\

\noindent For person $i$, let $T_{i0} = 0$ and $T_{i1}, T_{i2}, \ldots, T_{iN_i}$ represent recurrent event times. Here, $N_i$ counts events before $X_i = \min(C_i, D_i)$, where $C_i$ is censoring time and $D_i$ is CV death time. The hazard functions define the JFM:
\begin{align*}
r_i(t \mid \omega_i) &= \omega_i \exp\{\beta_1z_i\}r_0(t) = \omega_ir_i^*(t) \\
\lambda_i(t \mid \omega_i) &= \omega_i^{\alpha} \exp\{\beta_2z_i\}\lambda_0(t) = \omega_i^{\alpha}\lambda_i^*(t).
\end{align*}

\noindent For patient $i$, $r_i$ gives the hospitalization hazard, scaling with baseline intensity $r_0$ and frailty $\omega_i$. The CV death hazard $\lambda_i$ builds on baseline $\lambda_0$, with $\beta_1$, $\beta_2$ capturing treatment effects ($z_i$). Random effects $\omega_i$ follow a lognormal distribution (mean 0, variance $\theta$).\\

\noindent Parameter $\theta$ measures correlation strength - higher values indicate stronger within-patient correlations but also between-patient variability. The $\alpha$ parameter controls how CV hospitalizations link to  death:
- $\alpha = 0$: Events are independent
- $\alpha < 0$: Higher frailty increases recurrence but decreases terminal event risk
- $\alpha > 0$: Higher frailty increases both risks. A value of $\alpha =$ 1 was used in our simulations. \\

\noindent Once the 200 subjects from stage 1 have completed their 12 month follow-up, the interim analysis PP is calculated. Depending on the calculated PP, as described in Table 1, additional subject data for Stage 2 is simulated. The number of Stage 2 subjects can range from 0 (futility at IA) to 400 (Promising$\_$High). \\

 \noindent After data for the full trial has been generated, we compute the F-S statistic for cohort 2 and calculate the final test statistic, a weighted combination of the F-S Z-statistics for cohorts as described previously. Success for the trial is evaluated at the 1-sided alpha of 0.025.  Each scenario is simulated 10,000 times to estimate the design's operating characteristics. These are presented in Tables 3 and 4, each table detailing Scenarios 1 to 5 in Table 2.
 \section*{Simulation Results}

Table 3 shows the results of the simulation without any SSR (i.e., fixed design, with futility stop only). Without SSR, the design demonstrates strong type I error control (2.3\%) with effective futility stopping (68.2\%) under the null hypothesis. The alternative scenarios show adequate power (82.9 - 90.9\%) with moderate sample sizes (373 - 379 patients). However, power drops substantially in middling scenarios (47 - 52\%), suggesting potential vulnerability when true treatment effects are at the lower boundary of clinical significance or less. Futility stopping rates inversely correlate with treatment effect magnitude, ranging from 2.8\% in favorable scenarios to 19.2\% in middling scenarios, indicating appropriate interim decision-making sensitivity.

\begin{table}[h!]
\caption{Trial Operating Characteristics Without Sample Size Re-estimation}
\makebox[\textwidth]{
\begin{tabular}{lccc}
\hline
Scenario & Power (\%) & Average N & Futility Stop (\%) \\
\hline
Null & 2.3 & 261.5 & 68.2 \\
Alternative & 90.9 & 378.6 & 2.8 \\
Alternative2 & 73.4 & 384.3 & 7.9 \\
Middling & 63.6 & 377.0 & 11.4 \\
Middling2 & 47.0 & 348.6 & 19.2 \\
\hline
\end{tabular}}
\end{table}

With SSR implementation using an Adaptive Design, (Table 4), the type I error remains well controlled (1.9\%) while maintaining high futility stopping rates (69.1\%). This is due to the pre-specified weights used in the calculation of the test statistic at final analysis, which were originally proven by Cui, Hung, and Wang (1999) to prevent type I error inflation regardless of the rule for altering the stage 2 sample size. The adaptive design achieves improved power in alternative scenarios (88.7 - 94.5\%), though this comes at the cost of larger average sample sizes (437 - 444 patients). The proportion requiring maximum sample size ranges from 21 - 27\% in alternative scenarios. For middling scenarios, SSR provides modest power improvements (to 53 - 59\%), but frequently requires maximum sample size (33\% of trials), suggesting that while SSR helps mitigate power deficits for smaller treatment effects, it may not fully resolve them without substantial sample size increases. In this particular case, the maximum Sample Size of 600 was based on the Sponsor's resource constraints.
\begin{table}[h]
\caption{Trial Operating Characteristics With Sample Size Re-estimation}
\makebox[\textwidth]{
\begin{tabular}{lcccc}
\hline
Scenario & Power (\%) & Average N & Futility Stop (\%) & Max SSR (\%) \\
\hline
Null & 1.9 & 286.8 & 69.1 & 12.5 \\
Alternative & 94.5 & 437.0 & 2.5 & 21.3 \\
Alternative2 & 81.5 & 473.4 & 8.4 & 45.1 \\
Middling & 72.2 & 471.7 & 11.7 & 48.1 \\
Middling2 & 53.3 & 425.6 & 20.0 & 32.8 \\
\hline
\end{tabular}}
\end{table}

\section*{Discussion}
Adaptive sample size re-estimation designs offer several advantages, including improved trial efficiency, cost savings, increased flexibility, and enhanced statistical validity. While adaptive designs are well-studied and widely applicable for continuous, binary, and survival outcomes, they remain less common for complex endpoints. We propose a straightforward and practical approach for implementing sample size re-estimation in designs with complex hierarchical endpoints using the F-S statistic. This simulation-based approach uses the predetermined sample size and the Z-value from the interim analysis to estimate predictive power of trial success. Our approach enhances the applicability of adaptive designs to complex endpoints, unlocking their potential benefits for clinical trials, especially in the Cardiovascular space that employ the F-S statistic as the primary analysis method.

\pagebreak
\section*{References (Temp)}
1) Ajufo E, Nayak A, Mehra MR. Fallacies of Using the Win Ratio in Cardiovascular Trials: Challenges and Solutions. JACC Basic Transl Sci. 2023 Jun 26;8(6):720-727. 
\newline 
\newline
2) McCoy CE. Understanding the Use of Composite Endpoints in Clinical Trials. West J Emerg Med. 2018 Jul;19(4):631-634. doi: 10.5811/westjem.2018.4.38383. Epub 2018 Jun 4. PMID: 30013696; PMCID: PMC6040910.
\newline
\newline 
3) Finkelstein D, Schoenfeld D. Combining mortality and longitudinal measures in clinical trials. Stat Med. 1999;18:1341–1354.
\newline
\newline 
4) Marion, J et. al. (2024). Predictive Probabilities Made Simple: A Fast and Accurate Method for Clinical Trial Decision Making. arXiv preprint arXiv:2406.11406. 
\newline
\newline
5) Yu RX, Ganju J. Sample size formula for a win ratio endpoint. Statistics in Medicine. 2022; 41(6): 950–963.
\newline
\newline
6) Cui L, Hung HJ, Wang S-J. Modification of sample size in group sequential clinical trials. Biometrics. 1999;55:853-857.
\newline
\newline
7) Gao P, Ware JH, Mehta C. Sample size re-estimation for adaptive sequential design in clinical trials. J Biopharm Stat. 008;18(6):1184-96
\newline
\newline
8) Rogers et. al. Analysis of recurrent events with an associated informative dropout time: Application of the joint frailty model. Stat Med. 2016 Jan 10;35(13):2195–2205. 
\newline
\newline
9) Schoenfeld DA, Ramchandani R, Finkelstein DM. Designing a longitudinal clinical trial based on a composite endpoint: Sample size, monitoring, and adaptation. Stat Med. 2022 Oct 30;41(24):4745-4755. 
\newline
\newline
10) Pocock SJ et. al. The win ratio: a new approach to the analysis of composite endpoints in clinical trials based on clinical priorities. Eur Heart J. 2012 Jan; 33(2):176-82.
\newpage
\section*{Appendix: Using Conditional Power at Interim Analysis}

\noindent An alternative method to using approximate predictive power involves estimating the frequentist conditional power (CP) based on interim analysis. For the final analysis, let $n_2 = n - n_1$ represent the predetermined additional sample size for each arm in the second stage. The first 200 participants (from both arms combined) will be assigned to cohort 1, while the subsequent 200 planned participants (from both arms combined) will form cohort 2. Let
\[
w_i = \sqrt{\frac{n_i}{n}}, \quad i = 1, 2,
\]
be the weights that we will use to combine the F-S statistic $T_1$, obtained from the cohort 1 data, and the F-S statistic $T_2$, obtained from the cohort 2 data, such that the standardized test statistic for testing the null hypothesis at the final analysis is
\[
w_1 \frac{T_1}{s_1} + w_2 \frac{T_2}{s_2},
\]
where $s_j$ is the standard error of $T_j$. Then the CP, given $T_1 = t_1$, is
\[
CP(t_1, n_2) = P \left( w_1 \frac{t_1}{s_1} + w_2 \frac{T_2}{s_2} > z_\alpha \mid T_1 = t_1 \right).
\]

\noindent Since $T_1, T_2$ are independent and asymptotically normal, the CP can be evaluated as
\[
CP(t_1, n_2) = 1 - \Phi \left( z_\alpha - \frac{w_1 t_1}{s_1} - w_2 \frac{\mathbb{E}(T_2 / s_2)}{\sqrt{n_2}} \right),
\]
where $\mathbb{E}(T_2 / s_2)$ is estimated by repeatedly generating simulated responses for the $n_2$ subjects in cohort 2.

\subsubsection*{Sample Size Re-assessment Rule}

Based on the CP calculated above, once the 200 stage 1 participants complete their 12-month follow-up, the cohort 2 sample size may be increased from 200 participants to a maximum of 400 participants. The magnitude of this increase is determined using the promising zone approach proposed by Mehta and Pocock (2011). At the interim analysis, the data are categorized into Favorable, Unfavorable, and Promising Zones. A non-binding futility assessment based on CP can also be incorporated. If $CP(t_1, n_2)$ lies within the promising zone, the cohort 2 sample size is increased from $n_2 = 200$ upto a maximum of $n_2 = 400$, with the goal of achieving a CP of 90$\%$. Specifically, let $T_2$ represent the F-S statistic for the second cohort when the sample size increases from $n_2 = 200$ to $n_2 > 200$, and let $s_2$ denote its standard error. The augmented CP is then:
\[
CP(t_1, n_2) = 1 - \Phi \left( z_\alpha - w_1 \frac{t_1}{s_1} - w_2 \frac{\mathbb{E}(T_2 / s_2)}{\sqrt{n_2}} \right).
\]

\noindent Now let $n_{2\text{opt}}$ denote the value of $n_2$ at which $CP(t_1, n_{2,\text{opt}}) = 0.9$. (This involves repeated searches with different values of $n_2$.) Then
\[
n_2 = \min(400, n_{2,\text{opt}})
\]
and the corresponding CP is estimated as
\[
CP(t_1, n_2) = 1 - \Phi \left( z_\alpha - w_1 \frac{t_1}{s_1} - w_2 \frac{\mathbb{E}(T_2 / s_2)}{\sqrt{n_2}} \right).
\]
The estimates of $\mathbb{E}(T_2 / s_2)$ and $\mathbb{E}(T_1 / s_1)$ are obtained by repeatedly generating simulated responses for the $n_2$ or $n_2$ subjects in cohort 2. \\

\noindent During our evaluations, this method produced operating characteristics similar to those reported in the main paper using PP. However, it is significantly more numerically intensive, requiring substantial computing resources to optimize the Zones and CP thresholds. Using CP in this manner however allows for the precise calculation of Stage 2 sample sizes needed to achieve a CP of 90$\%$. This leads to a modest, but meaningful reduction in the average sample size required, while preserving statistical power. Another difference is that the CP assumes that the true value of the unknown parameter is equal to the observed treatment effect at IA. PP theoretically accounts for uncertainty in the true treatment effect by averaging over its posterior distribution. The PP calculation, as implemented, however is an approximation to the Bayesian PP using easily calculated quantities at the interim analysis. 

\end{document}